%% file: LPSolver.tex
\documentclass[AMA,STIX1COL,doublespace]{WileyNJD-v2}
\usepackage{moreverb}

\usepackage[caption=false,font=footnotesize]{subfig} 
\usepackage{xspace}

\DeclareMathOperator*{\maxi}{maximize}
\def \B {\ensuremath{\mathcal{B}}\xspace}

\newcommand{\Reals}[0]{\ensuremath{\mathbb{R}}}

\newcommand\BibTeX{{\rmfamily B\kern-.05em \textsc{i\kern-.025em b}\kern-.08em
T\kern-.1667em\lower.7ex\hbox{E}\kern-.125emX}}
\articletype{Article Type}%

\received{<day> <Month>, <year>}
\revised{<day> <Month>, <year>}
\accepted{<day> <Month>, <year>}

\begin{document}

\title{Simultaneous Solving of Batched Linear Programs on a GPU}

\author[1]{Amit Gurung*}

\author[2]{Rajarshi Ray}

\authormark{Amit Gurung \textsc{et al}}

\address[1]{\orgdiv{Department of Computer Science \& Engineering}, \orgname{National Institute of Technology Meghalaya}, \orgaddress{\state{Shillong}, \country{India}}}

\address[2]{\orgdiv{Department of Computer Science \& Engineering}, \orgname{National Institute of Technology Meghalaya}, \orgaddress{\state{Shillong}, \country{India}}}

\corres{*Amit Gurung, Department of Computer Science \& Engineering, National Institute of Technology Meghalaya. \email{amitgurung@nitm.ac.in}}

\presentaddress{Department of Computer Science \& Engineering, National Institute of Technology Meghalaya, Shillong - 793003, India}

\abstract[Abstract]{
Linear Programs (LPs) appear in a large number of applications and offloading them to a GPU is viable to gain performance. Existing work on offloading and solving an LP on a GPU suggests that there is performance gain generally on large sized LPs (typically 500 constraints, 500 variables and above). In order to gain performance from a GPU, for applications involving small to medium sized LPs, we propose batched solving of a large number of LPs in parallel. In this paper, we present the design and implementation of a batched LP solver in CUDA, keeping memory coalescent access, low CPU-GPU memory transfer latency and load balancing as the goals. The performance of the batched LP solver is compared against sequential solving in the CPU using the open source solver GLPK (GNU Linear Programming Kit) and the CPLEX solver from IBM. The evaluation on selected LP benchmarks from the Netlib repository displays a maximum speed-up of $95\times$ and $5\times$ with respect to CPLEX and GLPK solver respectively, for a batch of $1e5$ LPs. We demonstrate the application of our batched LP solver to enhance performance in the domain of state-space exploration of mathematical models of control systems design.
}


\maketitle

\input{Intro-Related.tex}

\input{application}

\input{LinearProgramming.tex}

\input{BatchedLP-GPU.tex}

\input{BoxLP-Application.tex}

\input{Conclusion.tex}

\section*{Acknowledgements} 
This work was supported by the National Institute of Technology Meghalaya, India and by the the DST-SERB, GoI under project grant No. YSS/2014/000623. The authors thank Santibrata Parida, for his help in experimental evaluations. 

\bibliography{mybibfile}
\section*{Author Biography}

\begin{biography}{\includegraphics[width=66pt,height=86pt]{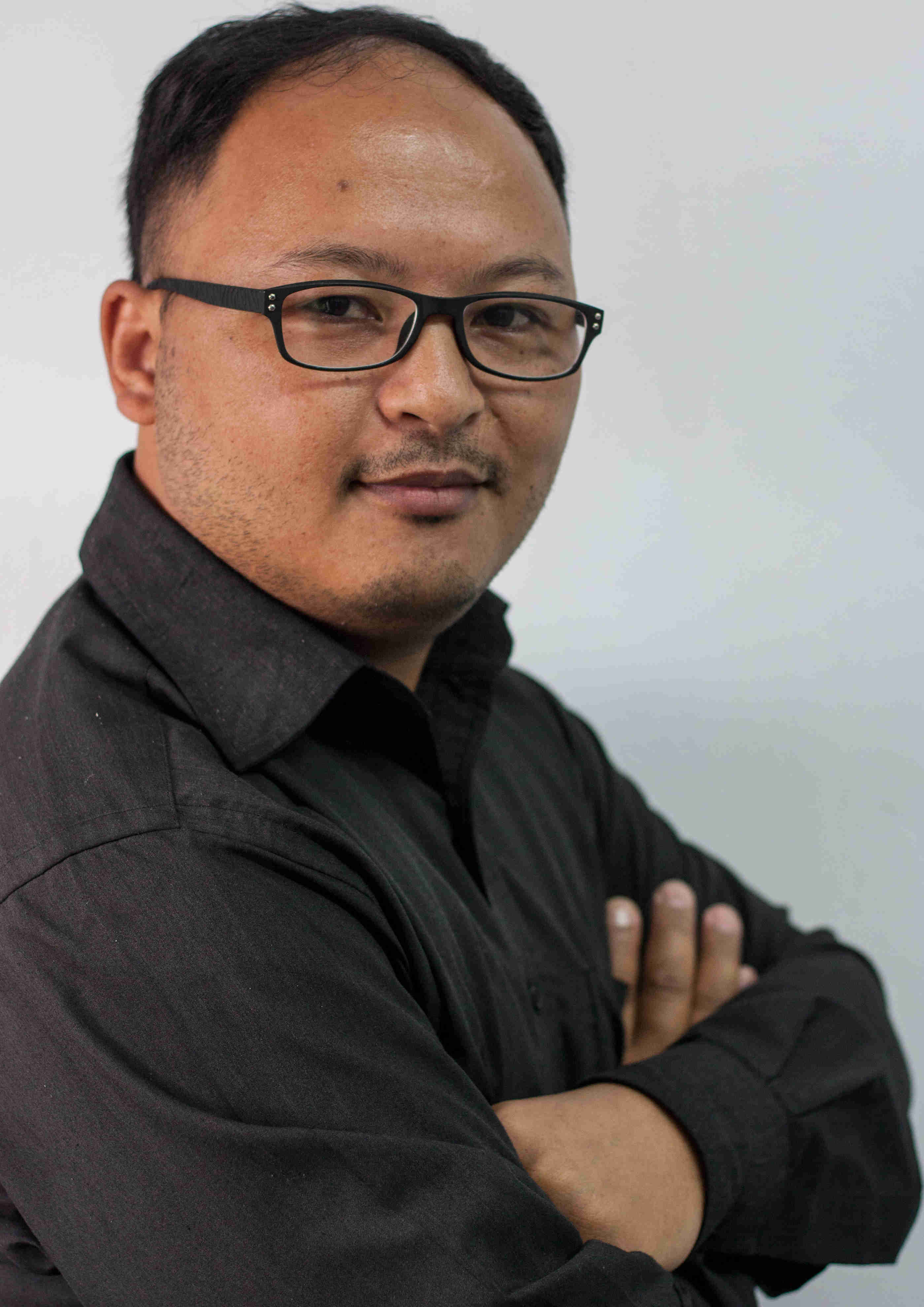}}{\textbf{Amit Gurung} is a Ph.D. student at the National Institute of Technology Meghalaya, India. He received an MCA (Master in Computer Applications) degree in 2010 and B.Sc. in Computer Science in 1999 from North Eastern Hill University, Shillong, India. His research includes High Performance Computing, Parallel Programming and Reachability Analysis of Hybrid Systems.}
\end{biography}

\begin{biography}{\includegraphics[width=66pt,height=86pt]{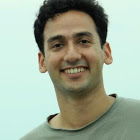}}{\textbf{Rajarshi Ray} is an Assistant Professor at the department of Computer Science and Engineering, National Institute of Technology Meghalaya. He received his B.Sc. in Computer Science from University of Pune in 2005 and M.Sc. in Computer Science from the Chennai Mathematical Institute in 2007. He received his Ph.D. in Computer Science from Verimag, University of Grenoble in 2012. His research interests are in formal methods in systems verification, cyber-physical systems and approximate computing.}
\end{biography}
\end{document}

%% file: Intro-Related.tex
\section{Introduction} \label{sec:Introduction}

Traditional CPU computations are now increasingly being carried out on CPU-GPU heterogeneous systems, by offloading data parallel tasks to a GPU for accelerating performance.  Some of the application domains where GPUs have been used for performance acceleration include medical image processing \cite{birk2014gpu,Yan2008}, weather research and forecasting (WRF) \cite{Michalakes2008}, Proteomics (to speed-up peptide spectrum matching \cite{Engineering2011}), signal processing for radio astronomy \cite{clark2012accelerating}, simulation of various physical and mechanical systems (using variants of Monte Carlo algorithm) \cite{karimi2011high,lim2015high} and large scale graph processing \cite{Shirahata2012}. However, drawing performance from a GPU requires insights on its architecture. Some of the common challenges of computing in a heterogeneous architecture are load balancing, efficient memory access and an effective mapping of computations in the SIMD paradigm of computing.

Linear Programming is a method of maximizing or minimizing a linear objective function subject to a set of linear constraints. Linear programs (LPs) appear extensively in a large number of application domains such as business process modeling to maximize profit, economics to design optimized demand-supply model \cite{chandra1988economic}, transport assignment \cite{munkres1957algorithms}, job scheduling \cite{drozdowski2006scheduling} and packets routing in computer networks, to name just some. Our work is motivated from applications that require solving a large number of small-size independent LPs. We propose a hybrid CPU-GPU LP solver which can solve a batch of many independent problems simultaneously. Our work assumes the setting that computations begin in a CPU where LPs are created, batched and then offloaded to a GPU for an accelerated solution. The solutions are transferred back to the CPU from the GPU for further processing. As an example, we show an application of our batched LP solver in model-based design of control systems. In model-based design, state-space exploration is a standard technique to analyze models of control systems. State of the art methods and tools for state-space exploration, heavily rely on solving many independent LPs \cite{FLGDCRLRGDM11,DBLP:conf/hvc/RayGDBBG15}. Moreover, the LPs are generally of small size (the size of an LP is the number of constraints $\times$ number of variables). We show that using our batched LP solver, the state-space exploration tools can improve the performance significantly. 

Prior work on solving an LP on a GPU and on multi-GPU architectures are many. The focus of almost all such works has been on methods to improve the performance of algorithms to solve one LP at a time. Performance gain is reported generally when offloading large LPs of size $500$ (500 constraints, 500 variables) and above \cite{lalami2011multi,bieling2010efficient,DBLP:conf/ipps/SpampinatoE09,ploskas2015efficient,meyer2011multi,lalami2011efficient}. It is seen that for small size LPs, the time spent in offloading the problems from CPU to GPU memory is more than the time gained with parallel execution in the GPU. Our work in this paper target applications that involves solving small to medium size LPs, but many of them. Although modern LP solvers like CPLEX \cite{cplex2009v12} and GLPK \cite{GLPK} are very efficient in solving small LPs, solving many of them one by one may consume considerable time.  Note that using any of the prior work to solve an LP at a time in GPU will not provide acceleration in such applications since they perform well only on large LPs. We show that with batched computation, performance acceleration can be achieved even for small LPs (e.g. LPs of size 5) for a considerably large \emph{batch size}, where \emph{batch size} refers to the number of LPs in a batch. We present a CUDA C/C++ implementation of a solver which implements the simplex method \cite{DantzigT97}, with an effort to keep coalescent memory accesses, efficient CPU-GPU memory transfer and an effective load balancing. To the best of our knowledge, this is the first work in the direction of batched LP solving on a GPU. Batched computations on a GPU to draw performance on small problem instances is recently adopted in \cite{haidar2015batched,abdelfattah2016performance,jhurani2015gemm,anderson2012predictive, villa2013accelerating, wainwright2013optimized}. The solver source and necessary instructions for repeatability evaluation can be found at \url{https://bitbucket.org/rajgurung777/simplexprojects}. We report solutions of LPs of dimension up to $511 \times 511$ ($511$ variables, $511$ constraints) with our solver. Beyond a sufficiently large batch size, our implementation shows significant gain in performance compared to solving them sequentially in the CPU using the GLPK library \cite{GLPK}, an open source LP solver and the CPLEX solver from IBM. The evaluation on selected LP benchmarks from the Netlib repository displays a maximum speed-up of $95\times$ and $5\times$ with respect to CPLEX and GLPK solver respectively, for a batch of $1e5$ LPs. In addition, we consider a special class of LPs having the feasible region as an hyper-rectangle and exploit the fact that these can be solved cheaply without using the simplex algorithm. We implement this special case LP solver as part of the solver.

The rest of the paper is organized as follows. Related works are discussed in Section \ref{sec:Related-Work}. A motivating application of batched solving of small LPs is discussed in Section \ref{sec:application}. In Section \ref{sec:LinearProgramming}, we discuss the simplex method that is needed to appreciate the rest of the paper. Section \ref{sub:Parallel-Algorithm} illustrates our CUDA implementation for solving batched LPs on a GPU, with memory coalescence, effective load balancing and efficient CPU-GPU memory transfer using CUDA streams. Section \ref{sec:AnalyseMultiLPs-GPU} presents an evaluation of our solver on batched solving of LP benchmarks in comparison to solving the same sequentially with GLPK and CPLEX. Section \ref{sec:XSpeedApply} demonstrates the performance enhancement in state-space exploration of models of control systems, on using our solver. We conclude in Section \ref{sec:Conclusion}.

\section{Related Work }\label{sec:Related-Work}

A multi-GPU implementation of the simplex algorithm in \cite{lalami2011multi} reports a speed-up of $2.93\times$ on dense randomly generated LPs of dimension $1000 \times 1000$. An average speed-up of $12.7\times$ has been reported for problems of dimension $8000 \times 8000$ or higher on a single GPU as compared to a sequential CPU implementation. An implementation of the revised simplex method using the OpenGL graphics library is reported in \cite{bieling2010efficient}. An average speed-up of $18\times$ is reported, compared to the GLPK library, for randomly generated dense problems of size $2700 \times 2700$ or higher. The paper also reports that the number of iterations in the simplex procedure is considerably less when the ``projected steepest-edge'' pivoting rule is used as against Dantzig's ``maximum reduced cost'' rule. A GPU implementation of the revised simplex algorithm is also reported in \cite{DBLP:conf/ipps/SpampinatoE09} with a speed-up of $2\times$ to $2.5\times$ in comparison to a serial ATLAS-based CPU implementation, for LPs of dimension $1400$ to $2000$. Automatically Tuned Linear Algebra Software (ATLAS\cite{whaley2011atlas}) is a library providing BLAS (Basic Linear Algebra Subprograms) APIs for C and Fortran. BLAS \cite{van2011blas} is a specification that prescribes routines for basic vector and matrix operations. A GPU-based implementation of the Revised Simplex algorithm and a Primal-Dual Exterior Point Simplex Algorithm (PDEPSA) is presented in \cite{ploskas2015efficient}. Pre-processing techniques on LPs have been proposed in order to gain performance. The LP solving algorithms use the steepest-edge pivoting rule which the authors claim to show the best performance in comparison to the other pivoting rules \cite{ploskas2014gpu}. Experimental results on Netlib \cite{dongarra2006netlib}, Kennington and M\`esz\`aros benchmarks shows that, on average the PDEPSA is $2.30\times$ and $32.25\times$ faster than the MATLAB's multi-threaded built-in function \textit{linprog}, which implements the interior point method, and a sequential CPU-based PDEPSA respectively. A CPU-GPU hybrid implementation of the Simplex algorithm is proposed in \cite{li2011gpu}. The conditional control instructions are executed in the CPU and the GPU execute the compute intensive tasks. Experimental result display a maximum speed-up of $120\times$ for LPs of size 7000. A single and a multi-GPU CUDA implementation of the standard simplex method with the steepest edge pivoting rule is proposed in \cite{meyer2011multi}. A comparison of their implementation with an open source solver, CLP is provided. It is shown empirically, that the performance of the revised simplex implementation (CLP) degrades with an increase in the density of the LP. However, the performance of their simplex implementation remains stable even when the LP density increases. Another implementation of the standard simplex method is presented in \cite{lalami2011efficient} for solving dense LP problems with double precision arithmetic. An average speed-up of $12.61\times$ is shown for LPs of size greater than 3000 as compared to the sequential CPU counterpart. We observe that almost all prior works report speed-up when the LP size is generally 500 or above. \cite{li2011gpu} is an exception which shows a marginal speed-up on LPs of size 100. 

%% file: application.tex
\section{Motivating Application}\label{sec:application}
In model-based design of control systems, a standard technique of analysis is to compute the state-space of the model using exploration algorithms. Properties of the control system such as safety and stability are analyzed by observing the computed state-space. In this section, we consider two open-source tools that perform state-space exploration of control systems with linear dynamics, namely SpaceEx \cite{FLGDCRLRGDM11} and  XSpeed \cite{DBLP:conf/hvc/RayGDBBG15}. These tools can analyze systems modeled using a mathematical formalism known as \emph{hybrid automaton} \cite{alur1993hybrid}. 
A conservative over-approximation of the exact state space is computed by both the tools. The state of the art state-space exploration algorithm in these tools compute the state-space as a union of convex sets, each having a symbolic representation, known as the support function representation \cite{GirardLG08}. The algorithm requires a conversion of these convex sets from its symbolic support function representation to concrete convex polytope representation, in order to have certain operations efficient. This conversion involves solving a number of linear programs. Moreover, the precision of the conversion and consequently, the precision of the computed state-space depends on the number of LPs solved. Table \ref{table:lpp} shows the number of LPs and its dimension that these tools solve for a fairly accurate state-space computation over a time horizon of just 100 seconds, on some standard control systems benchmarks. 

\begin{table}[!tbh]
\centering{}
\begin{tabular}{c  c  c}
\hline
Benchmark & LP Dimension & Number of LPs \\
\hline
Fourth Order Filtered Oscillator & 6 & $7.2e7$ \\
\hline
Eight Order Filtered Oscillator  & 10 & $2.0e8$\\
\hline
Helicopter Controller & 28 & $1.568e9$ \\
\hline
\end{tabular}
\caption{A large number of LP solving is required for a fairly accurate state-space computation.}
\label{table:lpp}
\end{table}

\begin{figure}[!htb]
\centering
\includegraphics[scale=0.5]{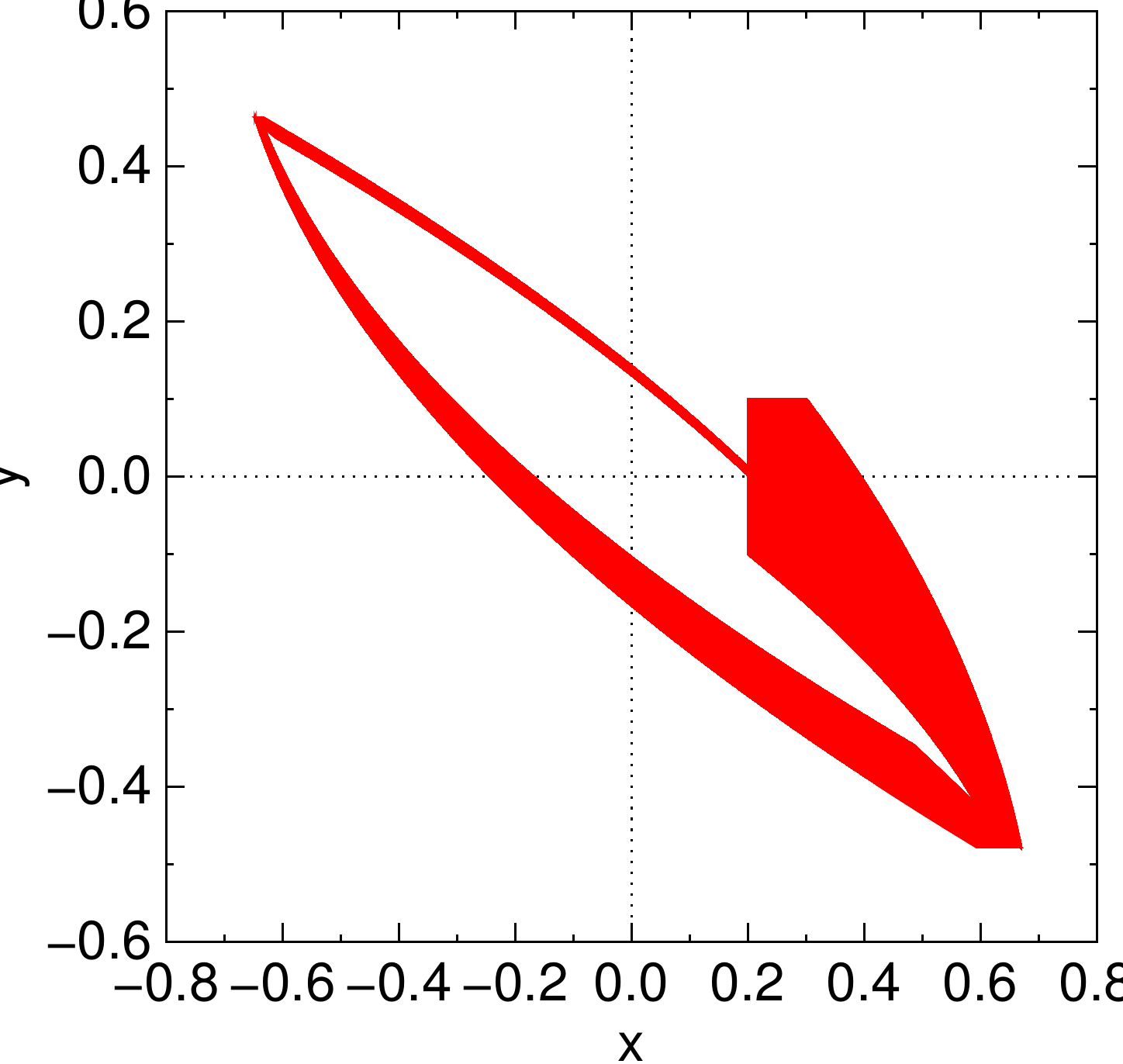}
\caption{State-space of a fourth order filtered oscillator model}
\label{fig:reach}
\end{figure}

We see that the number of LPs to be solved in the above examples is in the order of $1e9$ which cannot be solved in practical time limits even by the fast modern LP solvers like GLPK or CPLEX, when solved sequentially. Figure \ref{fig:reach} shows the computed state-space of a Filtered Oscillator model using the tool XSpeed. This computation required solving $7.2e7$ LPs. Note that although a solver like CPLEX take approximately $0.001$ seconds to solve an LP of dimension 6 in a modern CPU, it will take nearly 7 hours to solve $7.2e7$ LPs sequentially. Therefore, we believe that there is a need to accelerate applications where such bulk LP solving is necessary. A brief description of the control systems benchmarks of Table \ref{table:lpp} and the performance improvement when the tool XSpeed use our batched LP solver is reported in Section \ref{sec:XSpeedApply}.


%% file: LinearProgramming.tex
\section{Linear Programming} \label{sec:LinearProgramming}

A linear program in \emph{standard form} is maximizing an objective function under the given set of linear constraints, represented as follows:

\begin{equation}
maximize\qquad\sum_{j=1}^{n}c_{j}x_{j}\label{eq:Objective_Function}
\end{equation}

subject to the constraints

\begin{equation}
\sum_{j=1}^{n}a_{ij}x_{j}\le b_{i}\quad for\quad i=1,2,...,m\label{eq:constraints}
\end{equation}

and
\begin{equation}
x_{j}\ge0\quad for\quad j=1,2,...,n\label{eq:NonNegativeConstraints}
\end{equation}
In Expression (\ref{eq:Objective_Function}), $\sum_{j=1}^{n}c_{j}x_{j}$ is the objective function to be maximized and Inequality (\ref{eq:constraints}) shows the $m$ constraints over $n$ variables. Inequality (\ref{eq:NonNegativeConstraints}) shows the non-negativity constraints over $n$ variables. An LP in \emph{standard form} can be converted into \emph{slack form} by introducing $m$ additional \textbf{slack variables} ($x_{n+i}$), one for each inequality constraint, to convert it into an equality constraint, as shown below:

\begin{equation}
x_{n+i}=b_{i}-\sum_{j=1}^{n}a_{ij}x_{j},\:for\:i=1,...,m\label{eq:constraints-1}
\end{equation}
An algorithm that solves LP problems efficiently in practice is the \emph{simplex method} described in \cite{DantzigT97}. The variables on the left-hand side of the Equation (\ref{eq:constraints-1}) are referred as \textbf{basic variables} and those on the right-hand side are \textbf{non-basic variables}. The \emph{initial basic solution} of an LP is obtained by assigning its non-basic variables to zero. The \emph{initial basic solution} may not be always feasible (when one or more of the $b_i$'s are negative, resulting in the violation of the non-negativity constraint). For such LPs, the simplex method employs a two-phase algorithm. In the first phase, a new \textbf{auxiliary LP} is formed by having a new objective function $z$, which is the sum of the newly introduced \textbf{artificial variables.} The \textbf{simplex algorithm} is employed on this auxiliary LP and it is checked if the optimal solution to the objective function is zero. If a zero optimal is found then it implies that the original LP has a feasible solution and the simplex method initiates the second phase. In the second phase, the feasible slack form obtained from the first phase is considered and the original objective function is restored with appropriate substitutions and elimination of the artificial variables. The simplex algorithm is then employed to solve the LP.

Prior to the simplex method, many LP solvers apply preconditioning techniques such as a simple geometric mean scaling in combination with equilibration to reduce the condition number of the constraint matrix in order to decrease the computational effort for solving an LP \cite{tomlin1975scaling, larsson1993scaling, elble2012scaling, ploskas2015computational}. In this work, we do not apply any pre-conditioning on the LP for simplicity and use the simplex algorithm described in the following section.

\subsection{The Simplex Algorithm \label{sub:The-Simplex-Algorithm}}

The simplex algorithm is an iterative process of solving a LP. Each iteration of the simplex algorithm attempts to increase the value of the objective function by replacing one of the basic variables (also known as the \textbf{leaving variable}), by a non-basic variable (called the \textbf{entering variable}). The exchange of these two variables is obtained by a \emph{pivot operation}. The index of the leaving and the entering variables are called the pivot row and pivot column respectively. The simplex algorithm iterates on a tabular representation of the LP, called the \textbf{simplex tableau}. The simplex tableau stores the coefficients of the non-basic, slack and artificial variables in its rows. It contains auxiliary columns for storing intermediate computations. In our implementation, we consider a tableau of size $p \times q$, where $p=m+1$ and $q=n+\textit{sum of slack and artificial variables} + 2$. The ($m+1$)th row stores, the best solution to the objective function found until the last iteration, along with the coefficients of the non-basic variables in the objective function.

\begin{figure}[h]
\centering{}
   \includegraphics[scale=0.9]{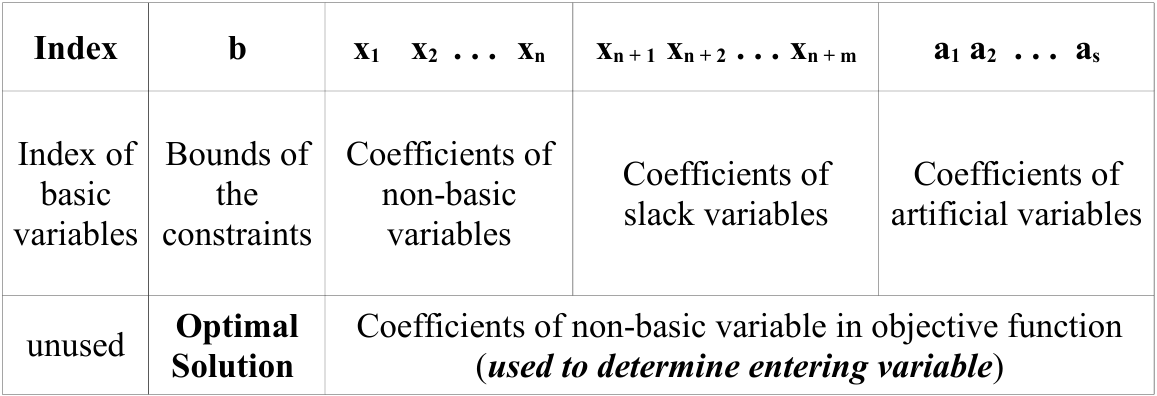}
\caption{Formation of the Simplex Tableau.}\label{fig:SimplexTableau}
\end{figure}

There are two auxiliary columns, the first column stores the index of the basic variables and the second stores $b_i$'s of inequality (\ref{eq:constraints}). Figure \ref{fig:SimplexTableau} shows a schematic of the simplex tableau.

\textbf{Step 1: Determine the entering variable. }

At each iteration, the algorithm identifies a new \emph{entering variable} from the non-basic variables. It is called an entering variable since it enters the set of basic variables. The choice of the entering variable is with the goal that increasing its value from 0 increases the objective function value. The index of the entering variable is referred to as the \emph{pivot column}. The most common rule for selecting an entering variable is by choosing the index $e$ of the maximum in the last row of the simplex tableau (excluding the current optimal solution). 

\textbf{Step 2: Determine the leaving variable. }

Once the pivot column is determined (say $e$), the algorithm identifies the row index with the minimum positive ratio $(b_i/-a_{i,e})$, say $\ell$, called the \emph{pivot row}. The variable $x_\ell$ is called the leaving variable because it leaves the set of basic variables. This ratio represents the extend to which the entering variable $x_e$ (in Step 1) can be increased without violating the constraints. 

\textbf{Step 3: Obtain the new improved value of the objective function. }

The algorithm then performs the \emph{pivot operation} which updates the simplex tableau such that the new set of basic variables are expressed as a linear combination of the non-basic ones, using substitution and rewriting. An improved value for the objective function is found after the pivot operation.

The above steps are iterated until the halt condition is reached. The halt condition is met when either the LP is found to be \emph{unbounded} or the \textbf{optimal solution} is found. An LP is unbounded when no new leaving variable can be computed, i.e., when the ratio ($b_i/ - a_{i,e}$) in Step 2 is either negative or undefined for all $i$. An optimal solution is obtained when no new entering variable can be found, i.e., the coefficients of the non-basic variables in the last row of the tableau are all negative values

%% file: BatchedLP-GPU.tex
\section{Simultaneous Solving of Batched LPs on a GPU} \label{sub:Parallel-Algorithm}
We present our CUDA implementation that solves batched LPs in parallel on a GPU. In this discussion, we shall refer a CPU by \emph{host} and a GPU by \emph{device}. The LP batching is performed on the host and transferred to the device. Our solver implementation assumes that all the LPs in a batch are of the same size. The batch size is adjustable, depending on the device memory size and LP size. Our batching routine considers the maximum batch size that can be accommodated in the device memory.

\subsection{CPU-GPU Memory Transfer and Load Balancing}\label{sec:memoryLoading}

First, we allocate device memory (global memory) from the host, that is required for creating a simplex tableau for the LPs in the batch. The maximum number of LPs that can be batched depends on the size of the device global memory in the device. The tableau for every LP in the batch is populated with all the coefficients and indices of the variables in the host side, before transferring to the device. To speed-up populating the tableau in the host, we initialize the tableau in parallel using OpenMP threads. Once initialized, the Simplex tableaux are copied from the host to the device memory (referred to as H2D-ST in Figure \ref{fig:DataOverlapWithKernel}). The LP batching routine is shown in Algorithm \ref{algo:batching}. The GPU kernel modifies the tableau to obtain solutions using the simplex method for every LP in the batch and copies back from the device to the host memory (referred to as D2H-res in Figure \ref{fig:DataOverlapWithKernel}). We discuss further on our CPU-GPU memory transfer using CUDA streams for efficiency in Section \ref{sub:streams}.

\begin{algorithm} [!htb]
  \caption{Batching Routine: $N$ -- the number of LP problems present in the data structure $listLP$. Computed results are returned in $R$}
  \begin{algorithmic}[1]
	\Procedure{batching}{$N, listLP, R$}
	   \State $gpuMem \gets getMemSize()$  \Comment{ get GPU's global memory size}
	   \State $lpSize \gets getLPSize()$  \Comment{ get memory requirement per LP}
	   \State $threadSize \gets getThreadSize()$	\Comment {computes the appropriate thread size based on LP dimension}
	   \State $batchSize= gpuMem \div lpSize$
		\If {$N > batchSize$} 
			\State $totBatch= ceil(N \div batchSize)$ \label{algo:batchSize}
			 \For{$i = \{0,...,(totBatch -1)\}$} 
			 	\State $start = i * batchSize$
			 	\If {$i==(totBatch-1)$}
			 		\State $end=N - 1$
			 	\Else
			 	 	\State $end = start + batchSize - 1$
			 	\EndIf
			 	\State $devLP \gets$ copy $listLP$ from index ($start$ to $end$)
				\State $batchKernel<<<batchSize, threadSize>>>(devLP, R)$	
			 \EndFor		
		\Else  
		 	\State $devLP \gets$ copy $listLP$ from index ($1$ to $N$)
			\State $batchKernel<<<batchSize, threadSize>>>(devLP, R)$
		\EndIf
	\EndProcedure
  \end{algorithmic}  \label{algo:batching}
\end{algorithm}

\paragraph{Load Balancing}

\begin{figure}[h]
\centering{}
   \includegraphics[scale=0.35]{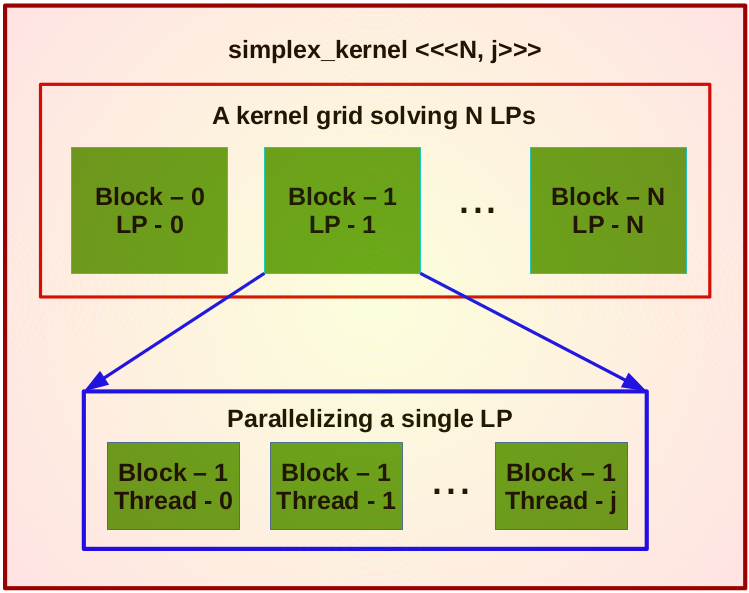}
\caption{Visualization of how threads are mapped to solve $N$ LPs in GPU. Each block is mapped to an LP and $j$ threads are assigned to parallelize a single LP.\label{fig:SimplexKernel}}
\end{figure}

We assign a CUDA block of threads to solve an LP in the batch. Since blocks are scheduled to Streaming Multiprocessors (SMs), this ensures that all SMs are busy when there are sufficiently large number of LPs to be solved in the batch. As CUDA blocks execute asynchronously, such a task division emulates solving many LPs independently in parallel. Moreover, each block is made to consist of $j$ ($\ge q$) threads, which is a multiple of $32$, as threads in GPU are scheduled and executed as warps. The block of threads is utilized in manipulating the simplex tableaux in parallel, introducing another level of parallelism. In Figure \ref{fig:SimplexKernel}, we show a block diagram of the logical threads in CUDA, of our GPU kernel.  

.
\subsection{Implementation of the Simplex Algorithm} \label{sub:SimplexAlgoImplement}

Finding the pivot column in \textbf{Step 1} of the simplex algorithm above requires to determine the index of the maximum value from the last row of the tableau. We parallelize \textbf{Step 1} by utilizing $n$ (out of $j$) threads in parallel to determine the pivot column using \textbf{parallel reduction} described in \cite{Harris2007}. A parallel reduction is a technique applied to achieve data parallelism in GPU when a single result (e.g. min, max) is to be computed from an array of data. We have implemented a parallel reduction by using two auxiliary arrays, one for storing the data and the other for storing the array indices of the corresponding data. The result of a parallel reduction provides us the maximum value in the data array and its corresponding index in the indices array. 

We also apply parallel reduction in \textbf{Step 2} by utilizing $m$ (out of $j$) threads in parallel to determine the pivot row ($m$ being the row-size of the simplex tableau). It involves finding a minimum positive value from an array of ratios (as described in Step 2 above) and therefore ratios which are not positive needs to be excluded from the minimum computation. This leads to a conditional statement in the parallel reduction algorithm and degrades the performance due to warp divergence. Even if we re-size the array to store only the positive values, the kernel still contains conditional statements to check the threads that need to process this smaller size array. To overcome performance degradation with conditional statements, we substituted a large positive number in place of ratios that are negative or undefined. This creates an array that is suitable for parallel reduction in our kernel implementation.

Data parallelism is also employed in the pivot operation in \textbf{Step 3}, involving substitution and re-writing, using the ($m-1$) threads (out of $j$ threads in the block).

\subsection{Coalescent Memory Access} \label{sub:memoryCoalescing}

In this section, we discuss our efforts on keeping a coalescent access to global memory to reduce performance loss due to cache misses. When threads in a warp access contiguous locations in the memory, the access is said to be coalescent. A coalescent memory access results in performance benefits due to an increased cache hit rate.

As discussed earlier, we use global memory to store the simplex tableaux of the LPs in a batch as described in Section \ref{sec:memoryLoading} (Since the global memory in a GPU is of the maximum size in the memory hierarchy, it can accommodate many tableaux). We store the simplex tableau in memory as a two-dimensional array. High level languages like C and C++ use the row-major order by default for representing a two-dimensional array in the memory. CUDA is an extension to C/C++ and also use the row-major order. The choice of row or column major order representation of two-dimensional arrays plays an important role in deciding the efficiency of the implementation, depending on whether the threads in a warp access the adjacent rows or adjacent columns of the array and what is the offset between the consecutive rows and columns.

We use the term \textit{column-operation}, when elements of all rows from a specific column are accesses simultaneously by each thread in a warp. If the array is in a row-major order, then this operation is not a coalesced memory access, as each thread access elements from the memory separated by the size equal to the column-width of the array. When elements of a specific row are accessed simultaneously by threads of a warp, we called this a \textit{row-operation}. Note that for a two dimensional array stored in row-major order, a row-operation is coalesced since each thread access data from contiguous locations in the memory.

We now show that in the simplex algorithm described above, there are more column-operations than row operations and thus, storing our data (i.e. simplex tableau) in a column-major order ensure more coalesced memory access in comparison to having a row-major storage.

\begin{figure}[h]
\centering{}
   \includegraphics[scale=1.0]{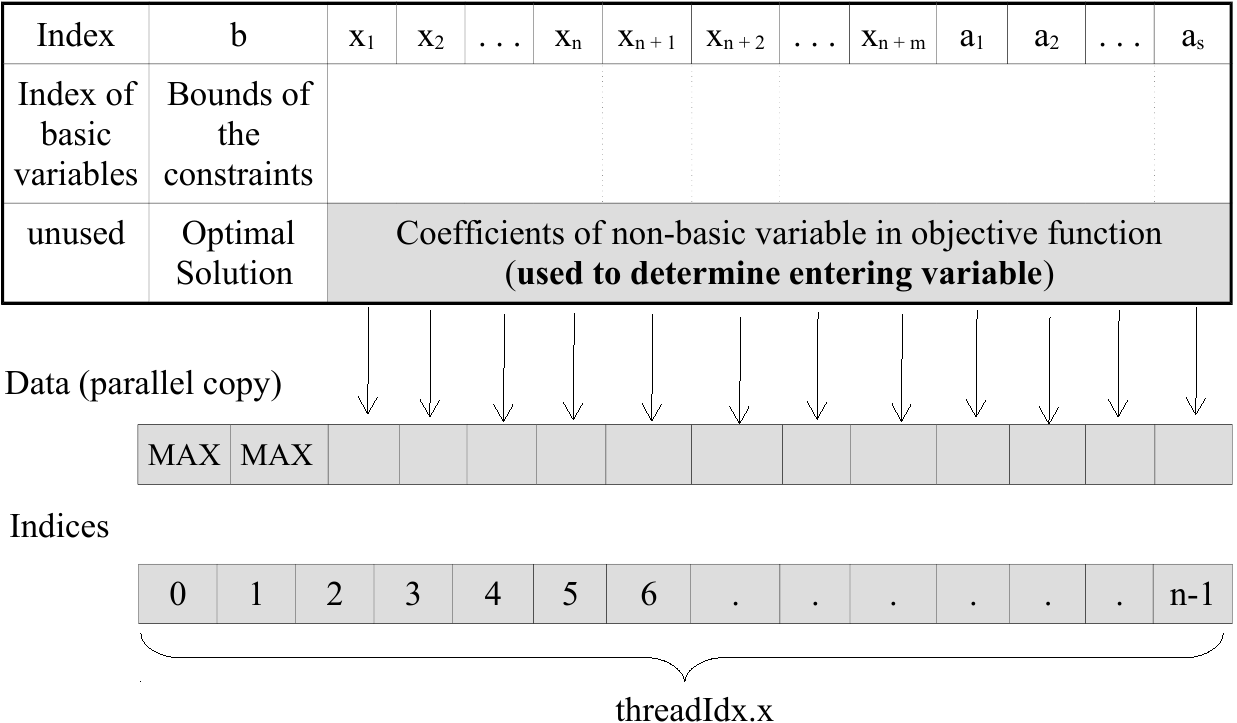}
\caption{Simplex Tableau along with two separate arrays, \textit{Data} to store the coefficients of the objective function and \textit{Indices} to keep track of the indices of the corresponding values in the \textit{Data} array.\label{fig:Step1}}
\end{figure}

\textbf{Step 1} of the simplex algorithm determines the entering variable (also known as the pivot column), which requires finding the index of the maximum positive coefficient from the last row. This requires a row-operation and as mentioned in Section \ref{sub:SimplexAlgoImplement}, we use parallel reduction using two auxiliary arrays, \textit{Data} and \textit{Indices} as shown in Figure \ref{fig:Step1}. Although accessing from the last row of the simplex tableau is not coalesced (due to our column-major ordering) but copying into the Data (and Indices) array is coalesced and so is the parallel reduction algorithm on the Data (and Indices) array. We use the technique of \emph{Parallel Reduction: Sequential Addressing} in \cite{Harris2007}, a technique that ensures coalesced memory access.

\begin{figure}[h]
\centering{}
   \includegraphics[scale=0.90]{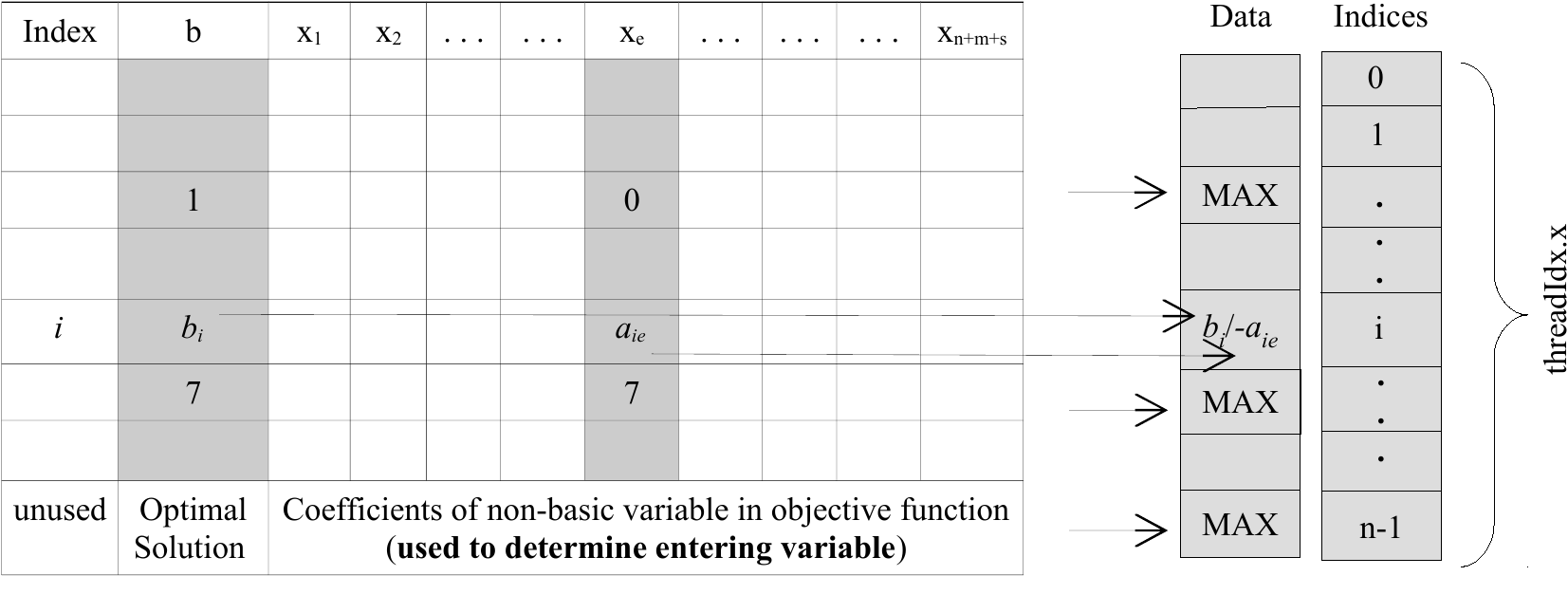}
\caption{Simplex Tableau along with two separate arrays, \textit{Data} to store the positive ratio and \textit{Indices} to keep track of the indices of the corresponding values in the \textit{Data} array. Ratios that reduces to negative or undefined are replaced by a large value denoted by MAX. \label{fig:Step2}}
\end{figure}

\textbf{Step 2} of the simplex algorithm determines the leaving variable (also called the pivot row) by computing the row index with the minimum positive ratio $(b_i/-a_{i,e})$, as described in Section \ref{sub:The-Simplex-Algorithm}. This requires two column-operations involving the access to all elements from columns $b$ and $x_e$ as shown in Figure \ref{fig:Step2}. To compute the row index with the minimum positive ratio, we use parallel reduction as described above in Section \ref{sub:SimplexAlgoImplement}. Our tableau being stored in a column-major order, access to columns $b$ and $x_e$ are both coalesced. The ratio and its corresponding indices (represented by the thread ID) are stored in the auxiliary arrays, \textit{Data} and \textit{Indices} which is also coalesced. Like in Step 1, we use the same technique of \emph{Parallel Reduction: Sequential Addressing} in \cite{Harris2007} for coalesced memory access.

\textbf{Step 3} performs the pivot operation that updates the elements of the simplex tableau and is the most expensive of the three steps. It first involves a non-coalescent row-operation which computes the new modified pivot row (denoted by the index $\ell$) as \{$NewPivotRow_{\ell} = OldPivotRow_{\ell} \div PE$\}, where PE is the element in cell at the intersection of the pivot row and the pivot column for that iteration, known as the pivot element. The modified row ($NewPivotRow_{\ell}$) is then substituted to update each element of all the rows of the simplex tableau, using the formula $NewRow_{ij} = OldRow_{ij} - PivotCol_{ie} * NewPivotRow_{\ell j}$ (see code Listing \ref{code:step3}). The elements of the pivot column are first stored in an array  named $PivotCol$ which is a column-operation, and so is coalesced, due to the column-major representation of the tableau. The crucial operation is updating each $j^{th}$ element for every $i^{th}$ row (except the pivot row $\ell$) of the simplex tableau, which requires a nested for-loop operation. We parallelize the outer for-loop that maps the rows of the simplex tableau. Our data being represented in a column-major order, parallel access to all rows for each element in the $j^{th}$ column of the inner for-loop is coalesced.

\begin{lstlisting}[language=c, caption={Code fragment for step 3 that updates the simplex tableau.}, label={code:step3}]

for (int i=0;i<rows;i++) { //Parallelized outer loop to map each i with the thread ID
  for (int j=0;j<cols;j++) {
    NewRow[i][j] = OldRow[i][j] - PivotCol[i] * NewPivotRow[l][j]; //l:pivot row index
  }
}
\end{lstlisting}

To verify the performance gain due to coalesced memory access, we experiment with \textbf{Step 3} which is the most expensive of the three steps in the simplex algorithm, by modifying it to have non-coalesced memory accesses. In the code Listing \ref{code:step3}, we interchange the inner for-loop with the outer loop (loop interchange, a common technique to improve cache performance \cite{hennessy2011computer}). This loop interchanging converts the Step 3 to have non-coalesced memory access since our simplex tableau is represented in a column-major order. Table \ref{fig:NonCoalescedResult} presents the experimental results to show the gain in performance when the accesses to memory are coalesced, as compared to having non-coalesced memory accesses. The results show a significant gain in performance on a Tesla K40c GPU, for LPs with the initial basic solution as feasible.

\begin{table*}[!htb]
\centering{}
   \includegraphics[scale=0.90]{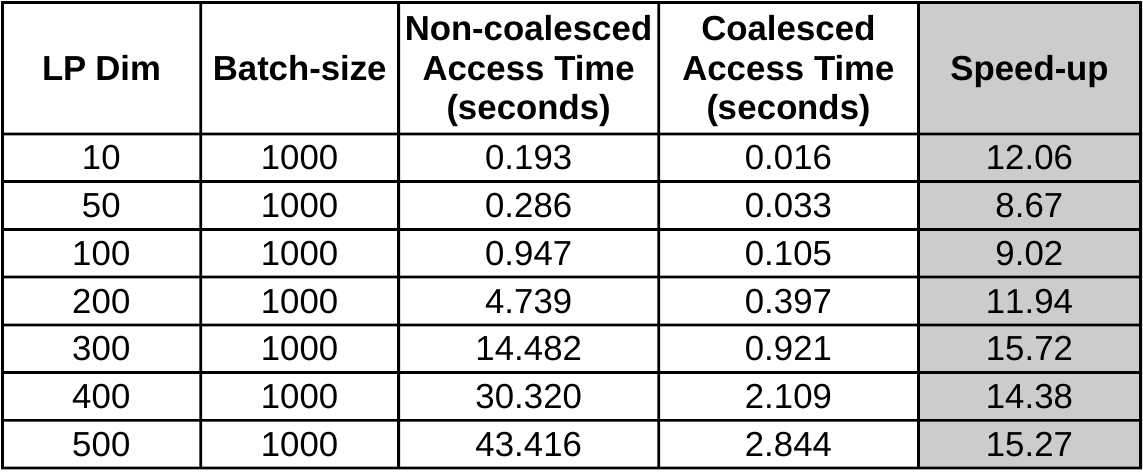}
\caption{Time taken to solve batched LP due to coalesced and non-coalesced memory access on a GPU, for LPs with the initial basic solution as feasible. \label{fig:NonCoalescedResult}}
\end{table*}


We observe that Step 1 has a row-operation, Step 2 has two column-operations and Step 3 has a row and a column operation each with a nested for-loop which can be expressed both in row as well as column operations. We see that there are more column operations than row operations. 

\subsection{Overlapping data transfer with kernel operations using CUDA Streams}\label{sub:streams}

The memory bandwidth of host-device data copy is a major bottleneck in CUDA applications. We use \textbf{nvprof} \cite{CudaGuide} to profile time for memory transfer and kernel operation of our implementation discussed above in Section \ref{sub:Parallel-Algorithm}. The result of profiling in a Tesla K40c GPU, for LPs with an initial basic solution as feasible, is reported in Table \ref{fig:profileResult}. We observe that, for a small batch size (e.g. $10$ in the Table \ref{fig:profileResult}), the memory copy operation takes a maximum of $5.75\%$ of the execution time, whereas for larger batch size, the memory copy operation takes $10\%$ to $15\%$ of the execution time.

\begin{table*}[!htbp]
\centering{}
   \includegraphics[scale=0.90]{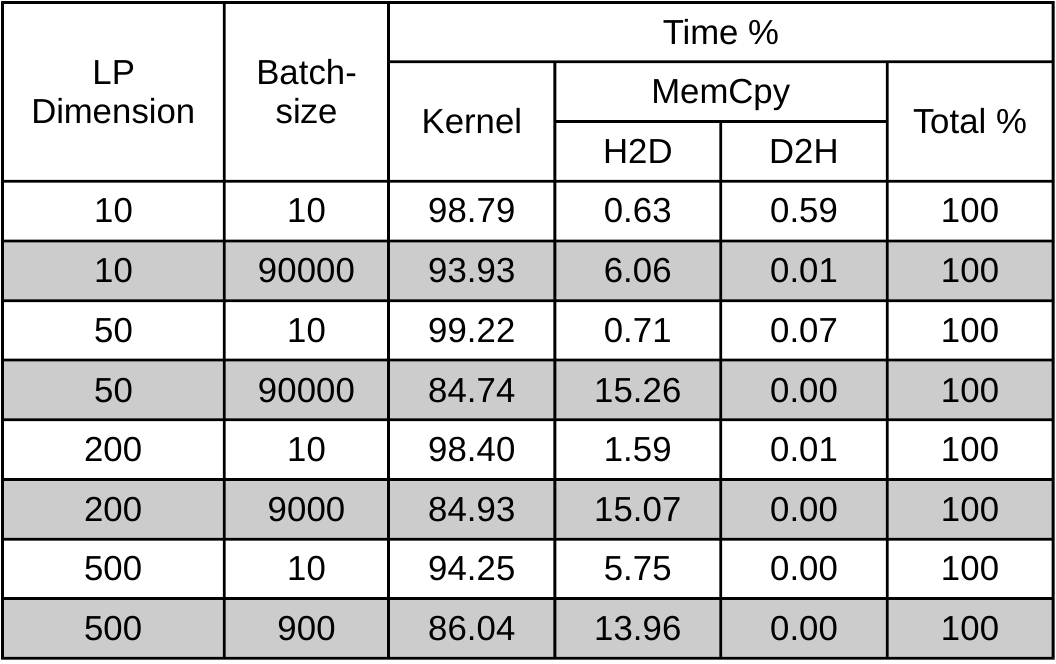}
\caption{Profiling report obtained using \textbf{nvprof} tool for LPs with an initial basic solution as feasible. H2D - stands for host to device and D2H indicates device to host memory copies respectively. \label{fig:profileResult}}
\end{table*}

A standard technique to improve the performance in CPU-GPU heterogeneous applications is by using CUDA streams. CUDA streams allow overlapping memory copy operation with kernel execution. A stream in CUDA consists of a sequence of operations executed on the device in the order in which they are issued by the host procedure. These operations can not only be executed in an interleaved manner, but also be executed concurrently in order to gain performance \cite{OverlapHarris2012}.

\begin{figure*}[!htb]
\centering{}
   \includegraphics[scale=0.55]{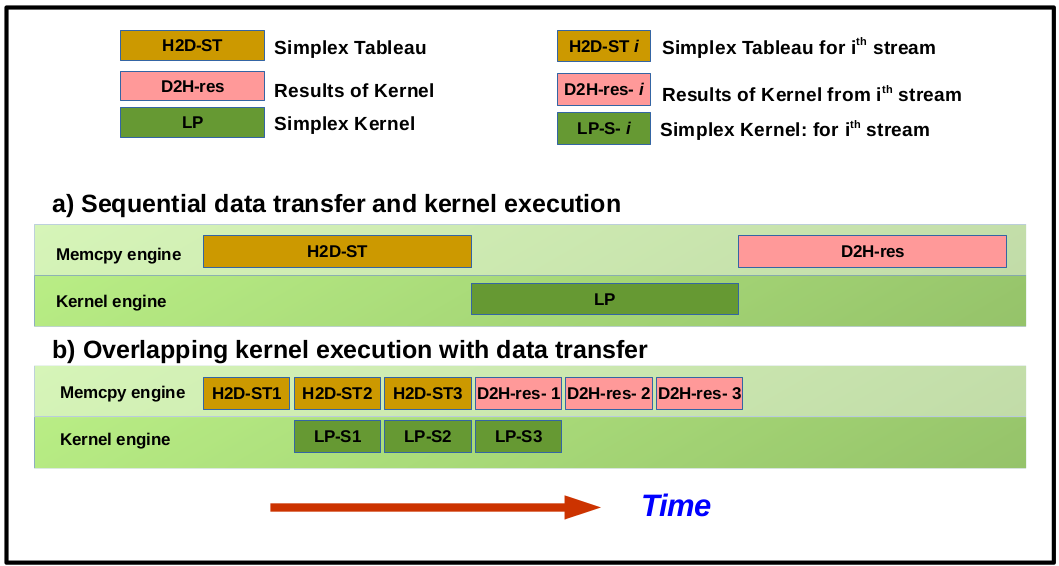}
\caption{Performance gain due to overlapping kernel execution with data transfer compared to sequential data transfer and kernel execution. The time required for host-to-device (H2D), device-to-host (D2H) and kernel execution are assumed to be the same.\label{fig:DataOverlapWithKernel}}
\end{figure*}

A GPU in general, has a separate kernel and a copy engine. All kernel operations are executed using the kernel engine and memory copy operations to and from the device are performed by the copy engine. However, some GPUs have two copy engines, one each for copying data from host to device and from device to host, for better performance. Figure \ref{fig:DataOverlapWithKernel} illustrates the overlapping of kernel executions with memory copy operation, when the GPU has one kernel and one copy engine. Streaming by batching similar operations causes more overlap of copies with kernel executions, as depicted in the figure. Adding all host-to-device copy to the CUDA streams followed by all kernel launches and device-to-host data copies, can result in a significant overlap of memory copy  operations with kernel executions, resulting in a performance gain. When there are two copy engines, looping the operations in the order of a host-to-device copy followed by kernel launch and device-to-host copy, for all streams may result in a better performance than the former method. For GPUs with compute capability 3.5 and above, both the methods result in the same performance due to the Hyper-Q \cite{HyperQEg} feature.

Although a large number of CUDA streams achieves more concurrency and interleaving among operations, it incurs stream creation overhead. The number of CUDA streams that gives optimal performance is found by experimentation. From our experimental observations, we conclude that with varying the batch size and the LP dimension, the optimal number of streams also varies. In this paper, we report results with $10$ streams for batch size more than $100$ LPs. We use a single stream when the batch size is less than $100$ (for LPs of any dimension).

\subsection{Limitations of the Implementation} \label{sub:implementation}
The memory required for an LP (i.e., a tableau) in our implementation can be approximately computed as:

\begin{equation}
Y=\{(m+1) \times cols \times dataSize +x\}
\label{eq:SingleLPSize}
\end{equation}

$\indent  \indent  \indent  \indent  \indent \indent \indent  \indent  \indent  \indent  \indent \indent \indent  \indent  \indent  \indent  \indent cols = (var+slack+arti+2)$ \\
$\indent  \indent  \indent  \indent  \indent \indent \indent  \indent  \indent  \indent  \indent \indent \indent \indent \indent  \indent \indent  \indent dataSize = sizeof(DType)$ \\
$\indent  \indent  \indent  \indent  \indent \indent \indent  \indent  \indent  \indent  \indent \indent \indent \indent \indent \indent  \indent  \indent x = 2 \times (cols \times dataSize)$ 

where $(m+1)$ and $cols$ are the sizes of rows and columns of the simplex tableau respectively, $var$ is the number of variables (dimension of the LP problem), $slack$ is the number of slack variables and $arti$ is the number of artificial variables in the given LP. The size of each LP is $Y$ bytes, where $DType$ is the data type of the LP coefficients and $x$ is the size of array used for performing parallel reduction operation, the number $2$ in the equation $x = 2 \times (cols \times dataSize)$ signify the use of two auxiliary arrays. The size of the $cols$ is described in Subsection \ref{sub:The-Simplex-Algorithm}. Thus, if $S$ is the size of total global memory (in bytes) available in the GPU, then our threshold limit or the number of LPs that can be solved at a time is determined by the equation $N = \lfloor\frac{S}{Y}\rfloor$. As the current limit on threads per block is $1024$ in CUDA, our implementation limits the size of LPs having \emph{initial basic solution as feasible} to $511 \times 511$. The size limit for LPs having \emph{initial basic solution as infeasible} is $340 \times 340$. This limit is derived from (\ref{eq:ThreadsPerBlock}). We intend to address this limitation in future work.
\begin{equation}
(var + slack + arti + 2) \le 1024 \label{eq:ThreadsPerBlock}
\end{equation} 

\subsection{Solving a Special Case of LP} \label{sec:HyperboxLP}
The feasible region of an LP given by its constraints defines a convex polytope. We observe that when the feasible region is a hyper-rectangle, which is a special case of a convex polytope, the LP can be solved cheaply. Equation (\ref{eq:HyperBoxEquation}) shows that maximizing the objective function is the sum of the results on $n$ dot products.

\begin{equation}\label{eq:HyperBoxEquation}
\maxi_{x \in \B} (\ell.x)=\sum_{i=1}^{n}\ell_{i}.h_{i},\text{where } h_{i} = 
\begin{cases}
a_i & \text{if $\ell_i<0$}\\
b_i & \text{otherwise}
\end{cases}
\end{equation}
where $\ell \in \Reals^n$ is the sampling directions over the given hyper-rectangle $\B = \{x \in \Reals^n | x\in [a_1, b_1] \times ... \times [a_n, b_n] \}$. 

An implementation for solving this special case of LPs is incorporated in our solver. In order to solve many LPs in parallel, we organize CUDA threads in a one-dimensional block of threads with each block used to solve an LP. Each block is made to consist of only 32 threads, the warp size. Within each block, we used only a single thread to perform the operations of the kernel. 
A preliminary introduction to this technique is introduced in the paper \cite{DBLP:conf/hvc/RayGDBBG15}.

\section{Evaluation} \label{sec:AnalyseMultiLPs-GPU}
\begin{figure*}[!htb]
	\centering
	\subfloat[5-Dimension \label{fig:5dLP}]
	{\includegraphics[scale=0.74]{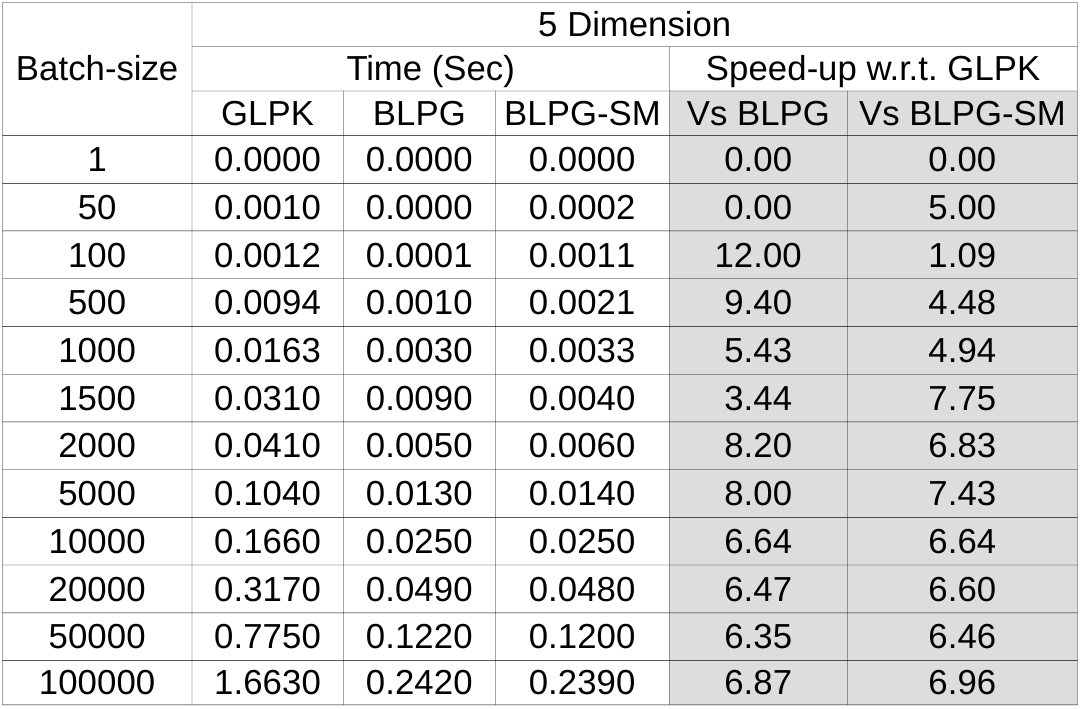}} \quad{}
	\subfloat[28-Dimension\label{fig:28dLP}]
	{\includegraphics[scale=0.73]{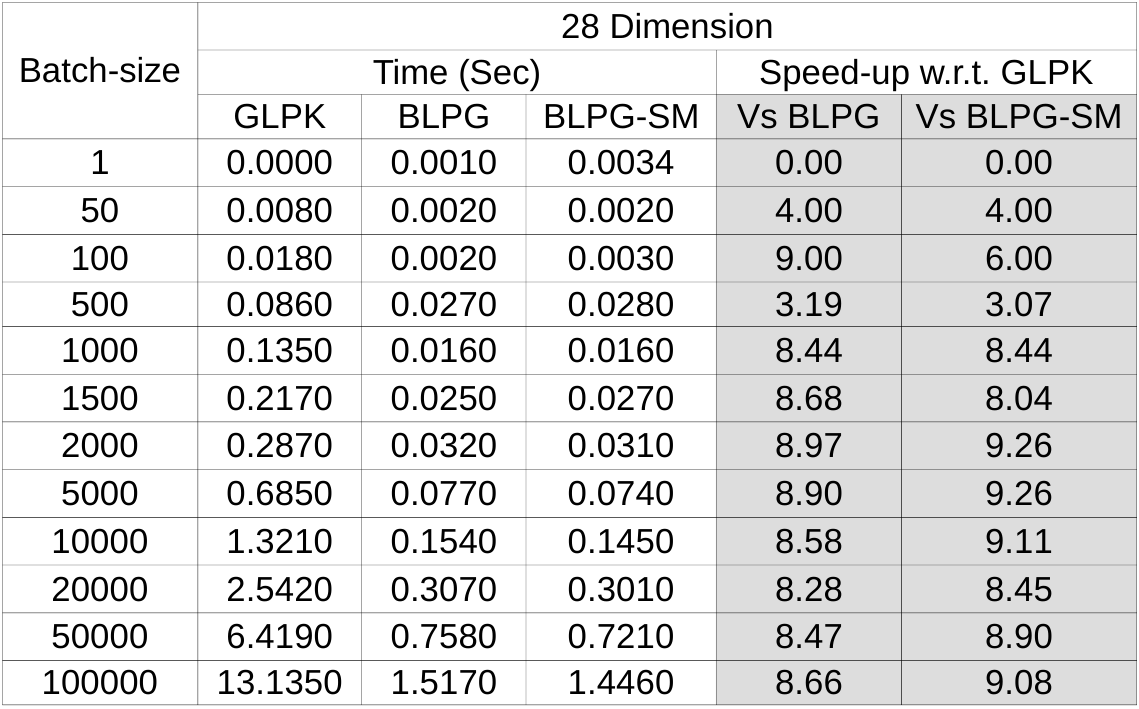}} 
	\hfill 
	\subfloat[50-Dimension\label{fig:50dLP}]
	{\includegraphics[scale=0.72]{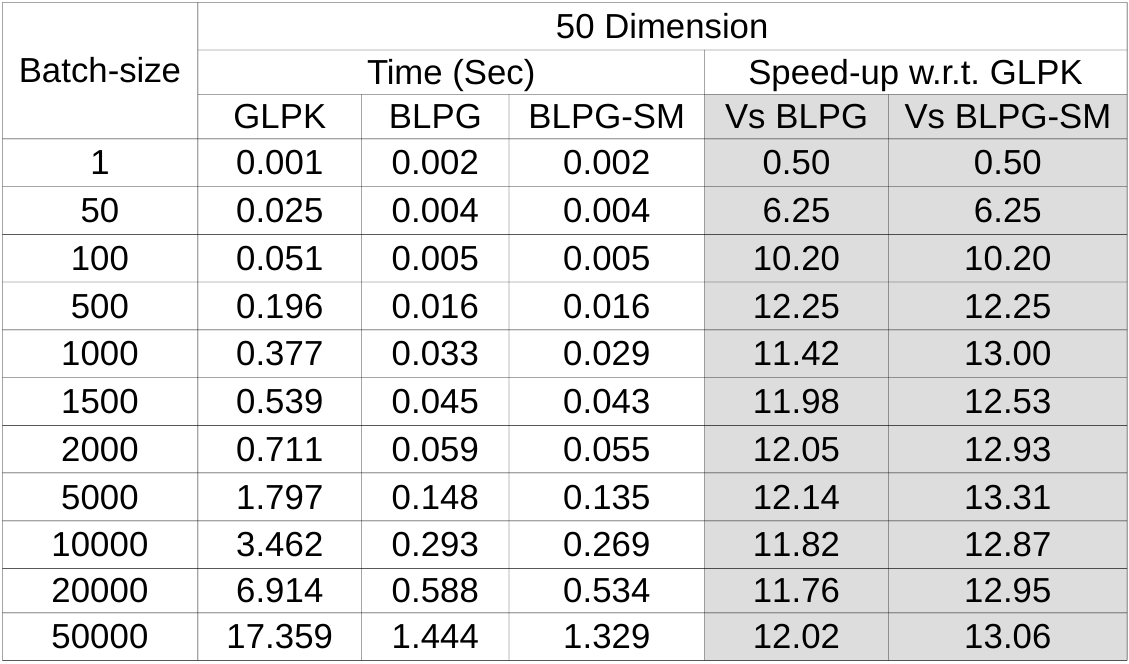}} \quad{}
	\subfloat[100-Dimension\label{fig:100dLP}]
	{\includegraphics[scale=0.72]{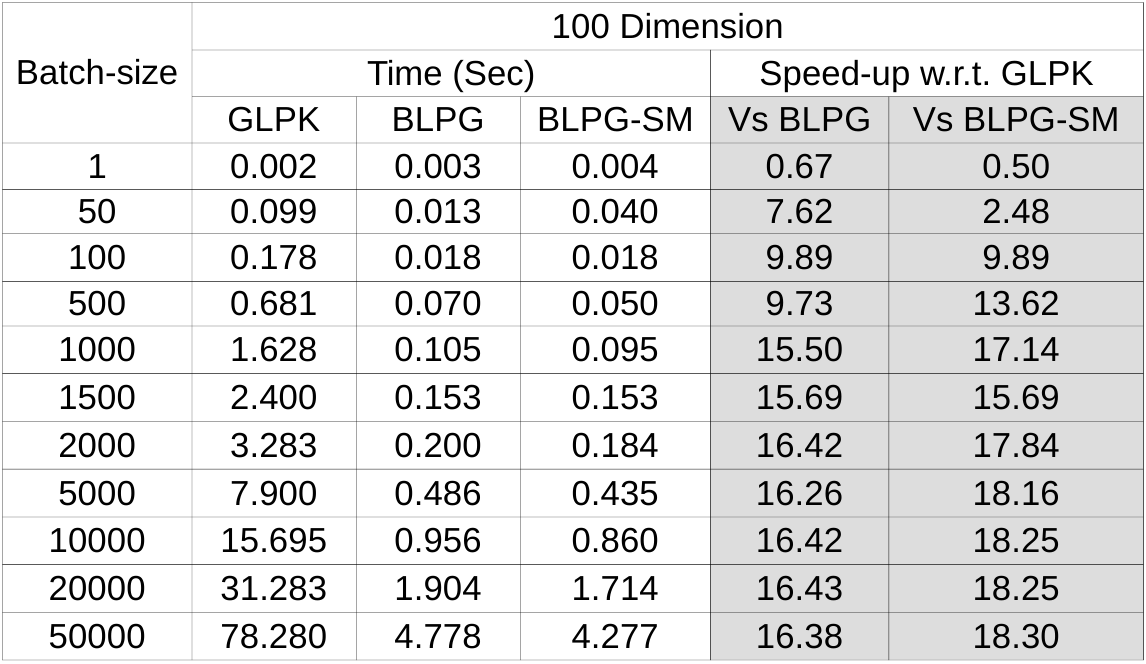}} 
	\hfill 
	\subfloat[300-Dimension\label{fig:300dLP}]
	{\includegraphics[scale=0.72]{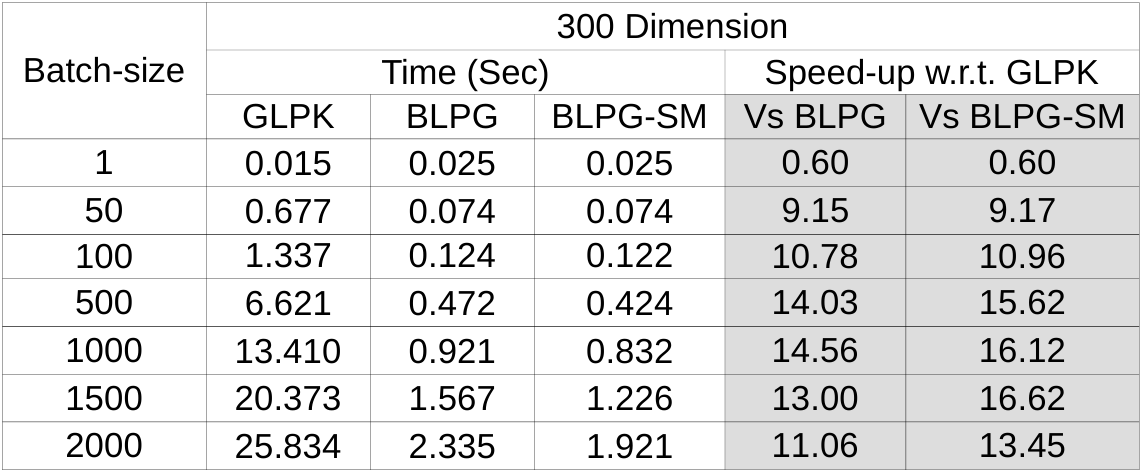}} \quad{}
	\subfloat[500-Dimension\label{fig:500dLP}]
	{\includegraphics[scale=0.72]{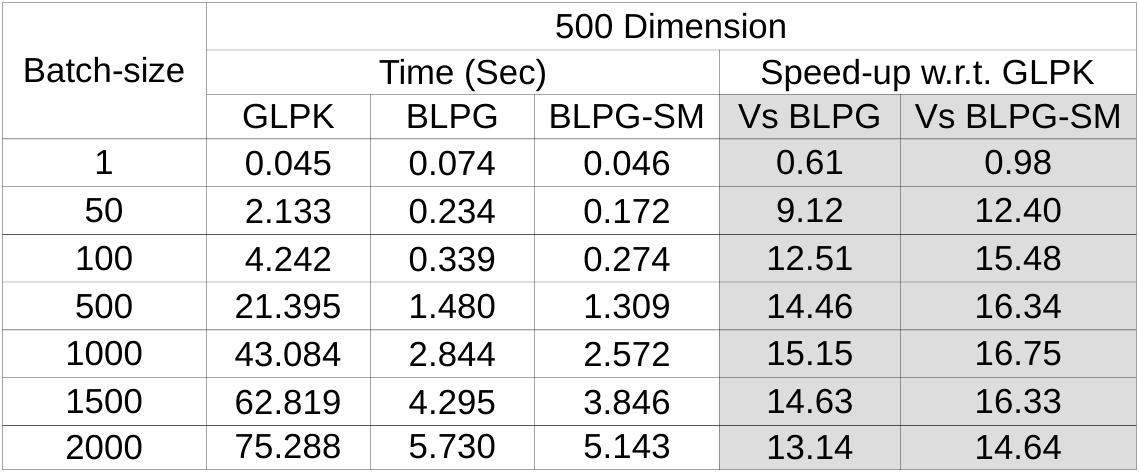}}
	
	\caption{Time taken to compute a batch of LPs of dimensions $5, 28, 50, 100, 300$ and $500$ respectively, for the type of LPs with \textbf{initial basic solution as feasible}.}	\label{fig:TimeLPplotting}
\end{figure*}

We evaluate our solver on two set of LPs. The first is a set of randomly constructed LPs. The LPs in this set are constructed by randomly selecting the coefficients of the constraint matrix $A$ from a range of $[1...1000]$, the bounds of the constraints $b$ from a range of $[1...1000]$ and the coefficients of the objective functions $c$ from the range of $[1...500]$ respectively. The second set of LPs are selected from the Netlib repository. On both the set of LPs, we evaluate the performance of our batched solver on varying batch sizes. In the text that follows, we refer to our solver on a GPU as BLPG, abbreviating Batched LP solver on a GPU. The solver using CUDA streams is referred as BLPG-SM. We perform our experiment in Intel Xeon E5-2670 v3 CPU, $2.30$ GHz, $12$ Core (without hyper-threading), $62$ GB RAM with Nvidia's Tesla K40c GPU. The reported performance is an average over $10$ runs. A performance comparison of our solver to GLPK is shown in Figure \ref{fig:TimeLPplotting}, for the first set of randomly generated LPs of various sizes and various batch sizes.  We observe a maximum speed-up of $16\times$ for LPs having the \emph{initial basic solution as feasible}, for a batch of 20k LPs of dimension 100. For the same type of LPs, we see a maximum speed-up of $18\times$, for a batch of $50k$ LPs of dimension 100, using BLPG-SM. We observe that for LPs of large size, BLPG performs better even with a few LPs in parallel (e.g., batch size=50 for a 500 dimensional LP). However, for small size LPs, BLPG out-performs GLPK only for larger batch sizes (e.g. a batch size of $100$ for a 5 dimensional LP).

Table \ref{tab:InfeasibleBasicSolution} shows a performance comparison of BLPG with GLPK on LPs having the initial basic solution as infeasible.
In-spite of the fact that BLPG executes the kernel twice due to the two-phase simplex algorithm as discussed in Section \ref{sec:LinearProgramming} (an extra overhead of data exchange between the two kernels), we observe a better performance. We gain a maximum speed-up of nearly $12\times$, for a batch of $10k$ LPs of dimension $200$.

\begin{table*}[!htb]
\centering{}
   \includegraphics[width=\textwidth]{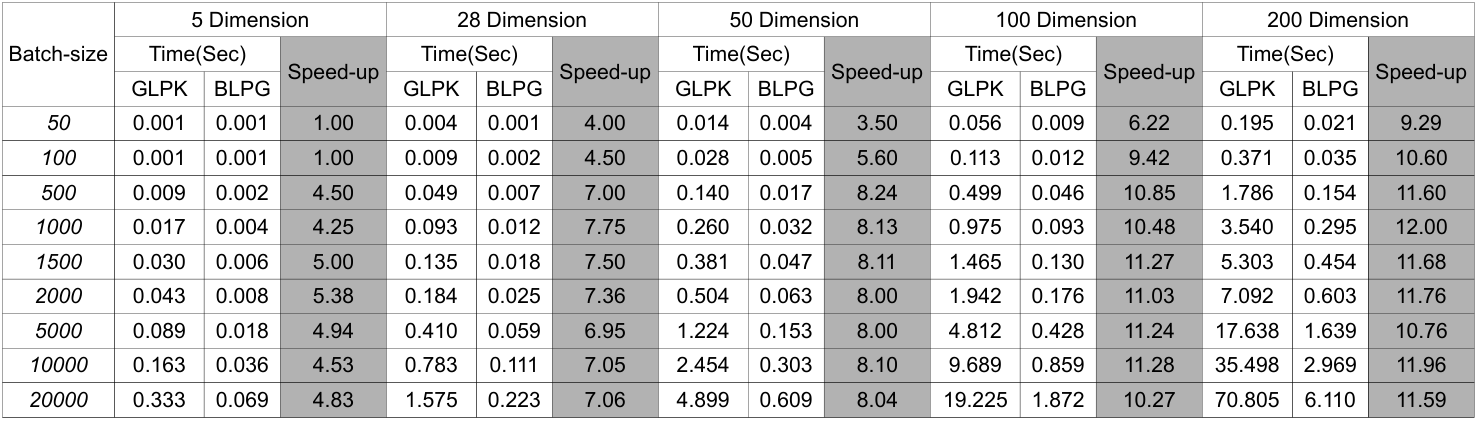}
\caption{Performance comparison between GLPK and BLPG for LPs with \textbf{initial basic solution as infeasible}}\label{tab:InfeasibleBasicSolution}
\end{table*}

On profiling BLPG-SM, we observe that for small sized LPs, the processing time of the kernel is much larger than the data transfer time, as shown in Table \ref{fig:profileResult}. As a result, the gain in performance due to overlapping data transfer with kernel execution is negligible. This is evident from the results in Figures \ref{fig:5dLP},\ref{fig:28dLP} and \ref{fig:50dLP}. As the LP size increases, the volume of data transfer also significantly increases. Hence, the operation of data transfer for all the streams (except the first) can be overlapped while the first kernel is in execution, thereby saving the time for data transfer in the rest of the streams. This results in $2-3\%$ performance gain for LPs of large sizes, as evident from Figure  \ref{fig:100dLP}, \ref{fig:300dLP} and \ref{fig:500dLP}.

\begin{table*}[!htb]
\centering{}
   \includegraphics[width=\textwidth]{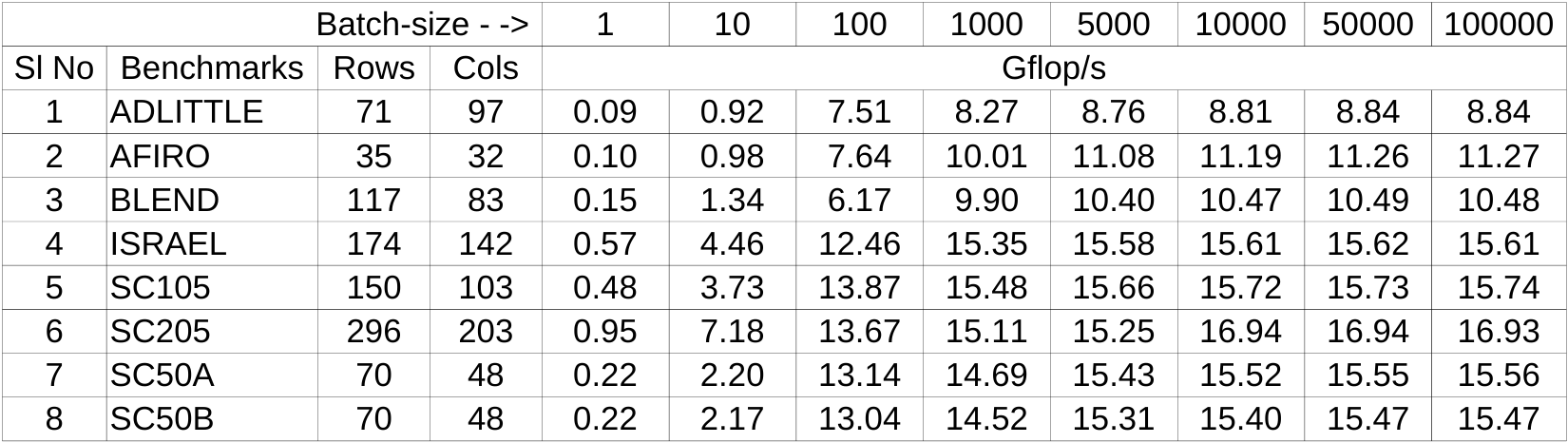}
\caption{Performance of batched LP solver on a GPU}\label{tab:GFlopSec}
\end{table*}

Table \ref{tab:BenchmarkLPs} shows a comparative evaluation of our solver to CPLEX and GLPK, on a set of selected LPs from the Netlib repository. The experimental platform is a $4$ Core Intel Xeon CPU E5-1607 v4, $3.10$ GHz, $63$ GB RAM with Nvidia's Tesla K40m GPU. The LP benchmarks in the Netlib repository are present in MPS format. We use the MATLAB's built-in function \textit{mpsread} to read the benchmarks and then convert them into the standard form denoted by Equations \ref{eq:Objective_Function} and \ref{eq:constraints}. The converted sizes of the benchmarks is shown in Table \ref{tab:GFlopSec} with column heading ``Rows'' indicating the number of converted constraints and ``Cols'' indicating the size of the benchmark. The table also shows the number of floating point operations per second (in Giga flops), giving an estimation of the floating point computations in the Simplex algorithm and the utilization of GPU by our proposed batched LP solver. We use the visual profiler (nvvp) available in the CUDA Toolkit \cite{nvidia2017toolkit}. The batched LP solver gives a maximum of $16.93$ Gflops/s for a batch size of $100K$ LPs on a Nvidia Tesla K40m GPU that has a theoretical peak of 1.43 Tflops/s, for double precision arithmetic.
\begin{table*} [!htbp]
	\centering{}
	\subfloat[Execution time of IBM CPLEX, GLPK and the proposed batched GPU implementation using double precision\label{fig:BenchmarkTime}]
	{\includegraphics[scale=1.1]{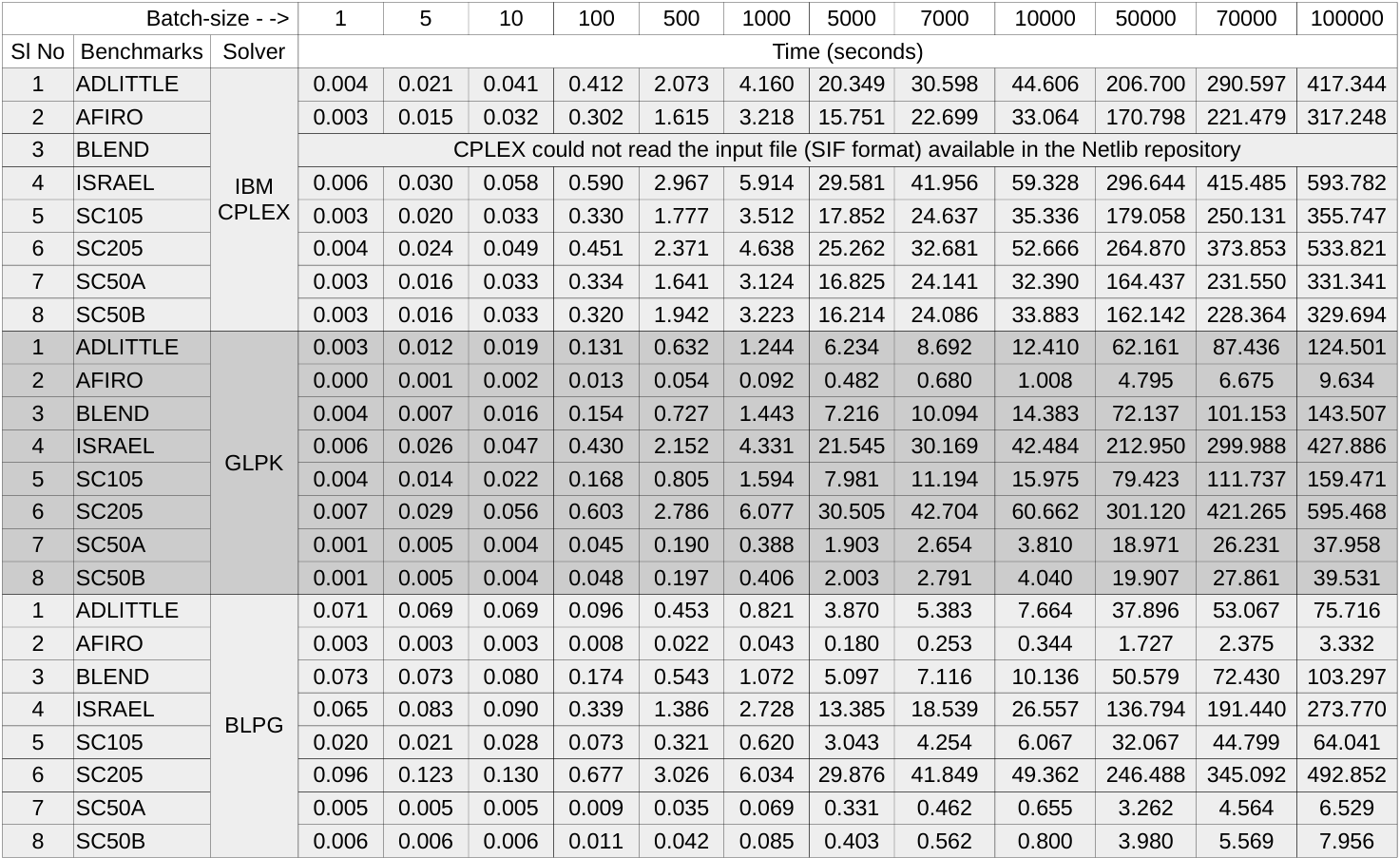}} 
	\hfill 
	\subfloat[Performance speed-up\label{fig:BenchmarkSpeedUp}]
	{\includegraphics[scale=1.1]{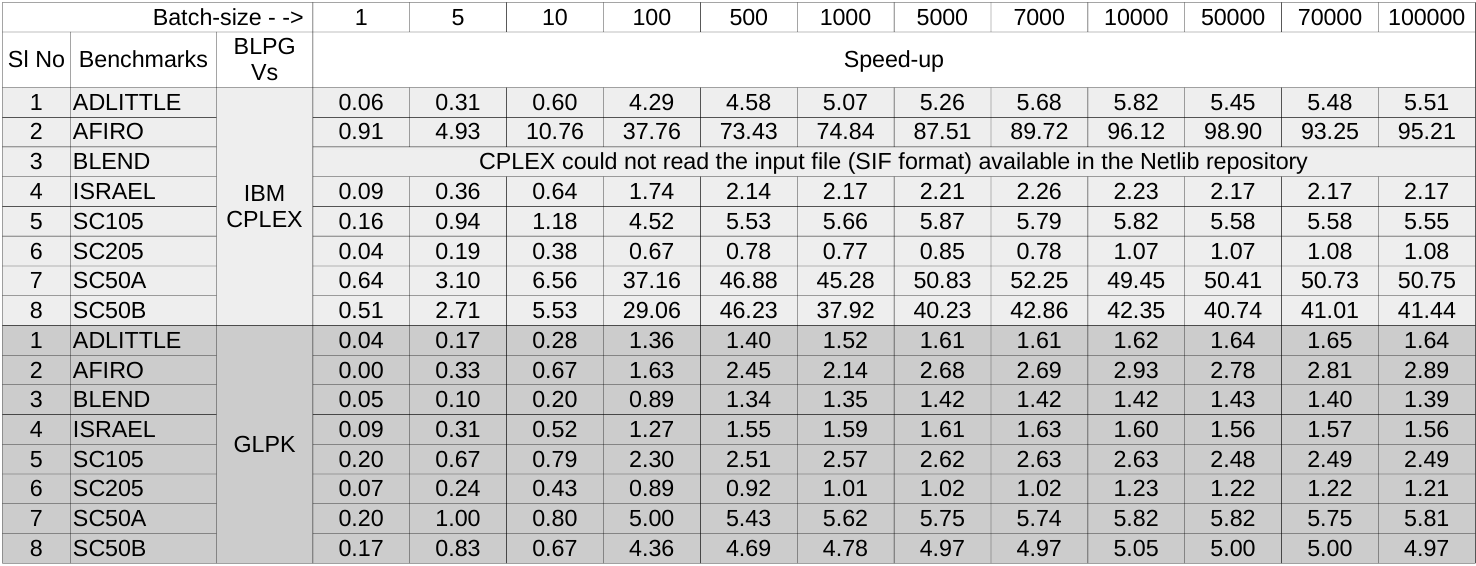}}	
\caption{Performance evaluation on selected benchmarks from Netlib}\label{tab:BenchmarkLPs}
\end{table*}

Due to the LP size limitation of BLPG discussed earlier in Section \ref{sub:implementation}, we choose benchmarks that satisfy this limitation. Table \ref{fig:BenchmarkTime} shows the execution time (in seconds) on the Netlib benchmarks by the three solvers and Table \ref{fig:BenchmarkSpeedUp} displays the performance comparison of the two LP solvers with that of BLPG. Note that Netlib benchmarks are highly sparse in nature and LP solvers such as IBM CPLEX and GLPK are optimized for sparse LPs. Our implementation is the original Simplex method proposed by Dantzig and does not have any optimization for sparse LPs. We observe that in some benchmarks such as \textbf{SC205} in Table \ref{tab:BenchmarkLPs} (and some others, not included in the table) CPLEX performs better than GLPK, whereas in other benchmark instances, GLPK outperforms CPLEX. Our proposed BLPG achieves a maximum of $95\times$ and nearly $6\times$ speed-up on the Netlib benchmarks w.r.t. CPLEX and GLPK respectively.

%% file: BoxLP-Application.tex
\section{Performance on Motivational Application}\label{sec:XSpeedApply}

In this section, we demonstrate the performance enhancement of state-space exploration of models of control systems, using our batched LP solver. As discussed in Section \ref{sec:application}, the state-space exploration routines in the state of the art tools requires solving a large number of LPs. We consider two benchmarks, the Helicopter controller and a Five dimensional dynamical system for its state-space computation using the tools \textsc{SpaceEx-LGG} and \textsc{XSpeed}. The Helicopter controller benchmark is a model of a twin-engined multi-purpose military helicopter with 8 continuous variable modeling the motion and 20 controller variables that governs the various controlling actions of the helicopter \cite{Skogestad:2005:MFC:1121635, FLGDCRLRGDM11}. The Five dimensional dynamical system benchmark is a model of a five dimensional linear continuous system as defined in \cite{Girard05}. We direct the reader to the paper \cite{Girard05} for details on the dynamics of the model.  

In order to show the impact of our solver BLPG, we evaluate the performance of the tools by solving the resulting LPs generated from the state-space exploration routines on these benchmarks. The LPs are sequentially solved using GLPK and in parallel using BLPG. The performance comparison is shown in Table \ref{tab:supGPU}, in an experimental setup of Intel Q9950 CPU, 2.84 Ghz, 4 Core (no hyper-threading), 8 GB RAM with a GeForce GTX 670 GPU. We observe a maximum speed-up of $12\times$ and $9\times$ in the tools performance with parallel solving of the LPs in BLPG w.r.t. sequential solving in GLPK. When compared to the tool \textsc{SpaceEx-LGG}, we observe a maximum of $54\times$ and $39\times$ speed-up in XSpeed using our solver. Note that the LPs generated by the tool on these benchmarks have the property that their feasible region is an hyper-rectangle. Therefore, BLPG solves these using the technique mentioned in Section \ref{sec:HyperboxLP}.

%

\begin{table*}[!htb]
\centering{}
 \includegraphics[scale=1.0]{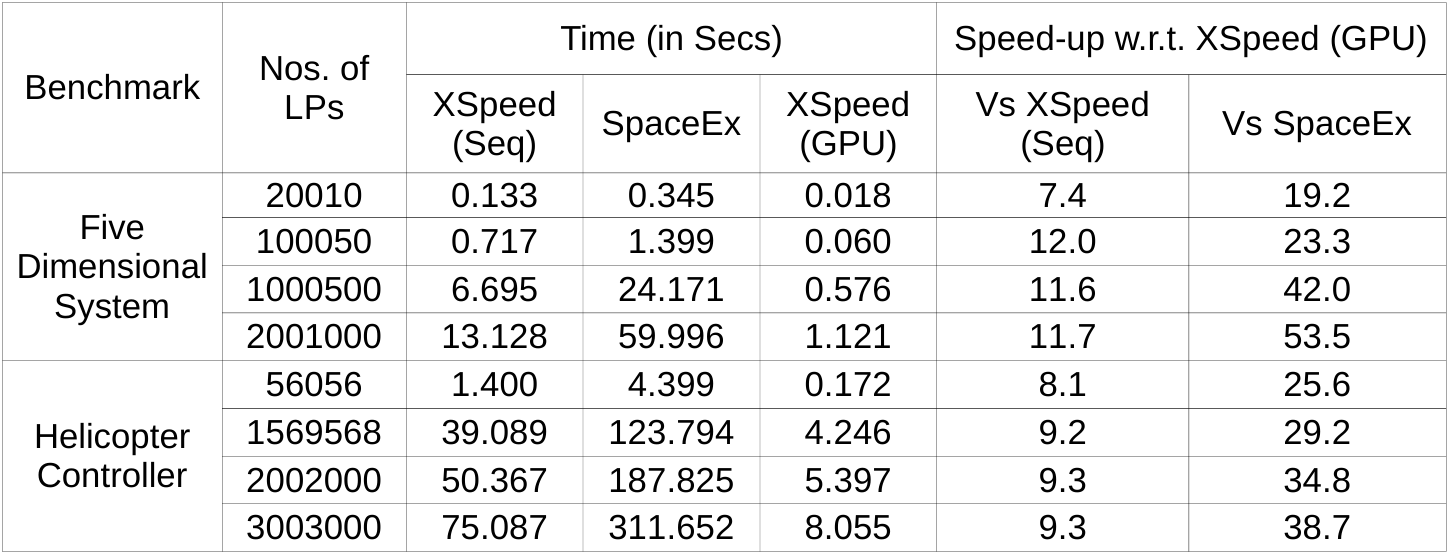}
\caption{Performance Speed-up in XSpeed using Hyperbox LP Solver}
\label{tab:supGPU}
\end{table*}


%% file: Conclusion.tex
\section{Conclusion }\label{sec:Conclusion}

Solving a linear program on a GPU for an accelerated performance on a CPU-GPU heterogeneous platform has been extensively studied. To the best of our knowledge, all such work report a performance gain only on linear programs of large size. We present a solver implemented in CUDA that can accelerate applications having to solve small to medium size LPs, but a large number of them. Our solver batches the LPs in an application and solves them in parallel on a GPU using the simplex algorithm. We report significant performance gain on benchmarks in comparison to solving them in CPU using GLPK and CPLEX solvers. We show the utility of our solver in an application of state-space exploration of models of control systems which involves solving many small to medium size LPs, by showing significant performance improvement.
